\documentclass [12pt]{article}
\usepackage{amsmath,amssymb,cite}
\usepackage{appendix}
\usepackage{feynmp}
\usepackage{float}
\usepackage[vcentermath,enableskew]{youngtab}
\usepackage{tabularx}
\usepackage[english]{babel}
\usepackage{graphicx}
\usepackage{indentfirst}
\usepackage{epsfig}
\usepackage{subfigure}
\usepackage{epstopdf}
\usepackage[]{hyperref}
\usepackage[section]{placeins}
\usepackage[stable]{footmisc}
\usepackage{appendix}
\usepackage{tabularx}
\usepackage[english]{babel}
\usepackage{graphicx}
\usepackage{indentfirst}
\usepackage{epsfig}
\usepackage{slashed}
\usepackage{fancyhdr}
\numberwithin{equation}{section}

\begin{document}

\title{
Quantum Gravity as a Multitrace Matrix Model}

\author{Badis Ydri $^{1,3}$\footnote{ydri@stp.dias.ie}, Cherine Soudani$^{2,1}$, Ahlam Rouag$^{1}$\\
$^{1}$ Department of Physics,  Badji Mokhtar Annaba University,\\
BP 12, 23000,Annaba, Algeria.\\
$^{2}$ Department of Physics,  Hamma Lakhdar El Oued University,\\
BP 789, 39000, El Oued, Algeria.\\
$^{3}$ The Abdus Salam International Centre for Theoretical Physics,\\
Strada Costiera 11, I-34014, Trieste, Italy.
}

\maketitle
\abstract{
We present a new model of quantum gravity as a theory of random geometries given explicitly in terms of a multitrace matrix model. This is a generalization of the usual  discretized random surfaces of 2D quantum gravity which works away from two dimensions and captures a large class of spaces admiting a finite spectral triple. These multitrace matrix models sustain emergent geometry as well as growing dimensions and topology change.
}

\section{Introduction}
Our understanding of the elementary particles and their fundamental interactions as well as our understanding of the fundamental laws governing the Universe as a whole were all reduced almost fully to geometry. This was started undoubtedly by Einstein revolution in the general theory of relativity a hundred years ago and the momentum of this continuous reduction by physicists of physics principles down to geometry principles is still very strong and much of their efforts is directed (directly or indirectly) to the completion of this reduction. 

In a word then, physics (not only gravity) is geometry. 

This is the basic story. But every story needs an "origin story" so to speak and the natural question which arises immediately: what is then the origin of geometry and how does it arise from nothing?. For obvious reasons, when we speak about geometry we are driven instinctively to thinking about the geometry of the spacetime manifold. Thus, the question more explicitly and more precisely is really what is the origin of the geometry of spacetime which is the theater in which all the cosmic play, i.e. the rest of physics, is unfolding and in the same time is the main actor of this cosmic play. 

One very plausible answer is given in terms of emergent geometry (classical but mostly quantum). Among the earliest, more drastic, and more imaginative original discussions of the idea that (quantum) geometry is an emergent concept are: causal sets of Sorkin et al. \cite{Bombelli:1987aa}, and loop quantum gravity and spin foams (see \cite{Thiemann:2007zz,Rovelli:2004} for an extensive list of references and a systematic discussion).

Another very powerful approach (or approaches) originates from superstring theory and M-theory. Their most important proposal of all (perhaps the most important proposal of all physics) is the gauge/gravity duality and in particular the AdS/CFT correspondence \cite{Maldacena:1997re}. This states that a maximally supersymmetric U(N) gauge theory in $(p+1)-$dimension in the 't Hooft limit $N\longrightarrow\infty$, $g_{\rm YM}^2\longrightarrow 0$ keeping $\lambda=g_{\rm YM}^2N$ fixed and large is equivalent to type II string theory in $10$ dimensions about a black p-brane solution which is formed as a bound state of $N$ coincident Dp-branes. Quantum gravity corrections in the string coupling $g_s$ are mapped to $1/N^2$ corrections on the gauge theory side and stringy corrections in the string length $l_s$ are mapped to $1/\lambda$ on the gauge theory side. 

Thus, a higher dimensional curved spacetime manifold emerges in this duality from a lower dimensional gauge theory in a flat spacetime manifold. The emerging extra spatial dimensions are described in the gauge theory by adjoint scalar fields given by $N\times N$ matrices. The extra dimensions emerges in the gauge theory precisely in the limit $N\longrightarrow \infty$ whereas strongly quantum gauge fields give rise to effective classical gravitational fields in the limit $\lambda\longrightarrow \infty$. See for example \cite{Horowitz:2006ct}. This duality provides therefore a very  concrete non-perturbative definition of quantum gravity and its quantum geometry in terms of gauge theory.

The other major proposal of string theory is M-theory and its M-(atrix) theory conjecture \cite{Banks:1996vh}. Indeed, the DLCQ (discrete light cone quantization) of M-theory should be described by the so-called M-(atrix) theory which corresponds to the above maximally supersymmetric U(N) gauge theory in $(0+1)-$dimension and therefore to the system of $N$ coincident D0-branes. M-(atrix) theory is thus a matrix quantum mechanics also known as the BFSS model. 

The compactification of the BFSS model on a circle (high temperature limit) gives the other powerful U(N) gauge theory in $(0+0)-$dimension known as the type IIB matrix model (a.k.a the IKKT model) \cite{Connes:1997cr,Ishibashi:1996xs}. The IKKT model provides a non-perturbative regularization of type IIB superstring theory in the Schild gauge \cite{Ishibashi:1996xs,Aoki:1998vn}.

In the IKKT and the BFSS matrix models the geometry of space is in a precise sense emergent given essentially by a spectral triple $({\cal A},\Delta,{\cal H})$, as in Connes' noncommutative geometry \cite{Connes:1996gi}, rather than in terms of a set of points. The algebra of functions  ${\cal A}$ on the underlying space is given by the algebra of $N\times N$ hermitian matrices, the Hilbert space ${\cal H}$ on which this algebra is represented is given by the adjoint representation of the gauge group U(N), whereas the Laplace operator is given in terms of the adjoint Casimir operator in the background solutions of the matrix models \cite{Sochichiu:2000ud}. This fuzzy or noncommutative geometry becomes a smooth manifold only in the large $N$ limit called in this setting the commutative limit.

Much of the emergent geometry scenarios discussed in the context of matrix field theory \cite{Ydri:2016dmy} is based on the BFSS and the IKKT matrix models and their lower dimensional analogues given by Yang-Mills matrix models. We may also consider mass deformations of these matrix models obtained by adding  mass terms such as the extent of space operator and/or the Myers term \cite{Myers:1999ps} to the Yang-Mills action, and as a consequence, the emergent geometry becomes a non-trivial space condensate. Indeed, in these cases the geometry appears dynamically as the system cools down in a phase transition from a pure algebra (called the matrix or Yang-Mills phase) to a background geometry given typically by some fuzzy space such as fuzzy ${\bf CP}^n$ \cite{Balachandran:2001dd}. 

The emergent noncommutative fuzzy geometry can be exhibited even further by expanding the BFSS and IKKT matrix gauge theories around the background solution in the geometric phase. 

The original adjoint scalar fields $X_a$ of the U(N) gauge theory are seen to split into a genuine gauge field on the background geometry plus a bunch of massive normal scalar fields. See for example \cite{Ydri:2016kua,Ydri:2016osu} and references therein. Furthermore, it is found that the background geometry is generically stable only in the region of the parameter space where the mass of these normal scalar fields is very large. An exception is perhaps the particular fuzzy sphere considered recently in \cite{Ydri:2016osu} where the background emergent geometry is observed to be completely stable for all values of the mass.  

So much for emergent space. But what about emergent time?

This is a far more complex but also a far more important question since an emergent time is directly and intimately related to the origin of the Universe \cite{Seiberg:2006wf}. A remarkable development on this front was given by the matrix simulations of the Lorentzian IKKT matrix gauge model \cite{Kim:2011cr}. The Lorentzian model as opposed to the Euclidean version does not suffer from the sign problem whereas the bosonic matrices $X_{\mu}$ give a phase factor $\exp(iS_b)$ in the path integral. By integrating out the scale factor of the bosonic matrices first \cite{Krauth:1998xh} we get then the constraint $S_b=0$. The model requires also an appropriate regularization before it can be accessed in the usual Monte Carlo importance sampling by putting cutoffs on the extents of space and time given respectively by the traces ${\rm Tr}X_i^2$ and ${\rm Tr}X_0^2$ which are seen to diverge otherwise. 

It is found that a large $N$ scaling limit exists in which the cutoffs can be removed and the theory is seen to depend only on a single parameter given by the scale factor which can be identified as the string scale. Furthermore, it is observed that the eigenvalue distribution of the matrix $X_0$ representing time has an infinite extent in the large $N$ limit which is a property traced to supersymmetry since in the bosonic model the  extent of the eigenvalue distribution of the time matrix $X_0$ is found to be finite. Also it is observed that the theory has an SO(9) Lorentz symmetry only until a critical time after which this symmetry gets spontaneously broken down to SO(3). This corresponds to the fact that only three directions $X_i$ start to expand rapidly at the critical time whereas the other six gets shrunken down. The nature of the spontaneous symmetry breaking in the Lorentzian model is due principally to the noncommutativity of space and therefore it is very different compared to the Euclidean case where it is caused by the phase of the Pfaffian. For more detail on this exciting results see \cite{Kim:2012mw,Kim:2011ts,Nishimura:2012xs}.

Another powerful approach to the emergence of time and spacetime is Lorentzian causal dynamical triangulation \cite{Ambjorn:2005qt,Ambjorn:2006hu,Ambjorn:2007jv}. In this approach spacetime is built out of four-simplices (generalization of two-simplices, i.e. triangles, to four dimensions) which are equipped with a flat Minkowski metric. The causality requirement singles out globally hyperbolic manifolds which admit a global proper-time foliation structure and as a consequence Wick rotation to Euclidean is meaningful. The Hilbert-Einstein action is given in this discrete setting by the Regge action \cite{Regge:1961px}. The path integral is obtained as the sum over the set of all causal triangulations weighted with the Regge action. The parameters of the model are Newton’s gravitational constant $G$ and the cosmological constant $\Lambda$ which appear as the parameters $K_0$ and $K_4$ in the Regge action. Also the model depends on two more parameters given by the lengths of time-like and spatial-like links $a_t$ and $a_s$ respectively. We have $a_t^2=\alpha a_s^2$ where the asymmetry factor $\alpha< 0$ appears as a parameter $\Delta$ in the Regge action.

\begin{figure}[H]
\begin{center}
\subfigure[Landau-Lifshitz scalar field theory.]
{
\includegraphics[width=8.0cm,angle=0]{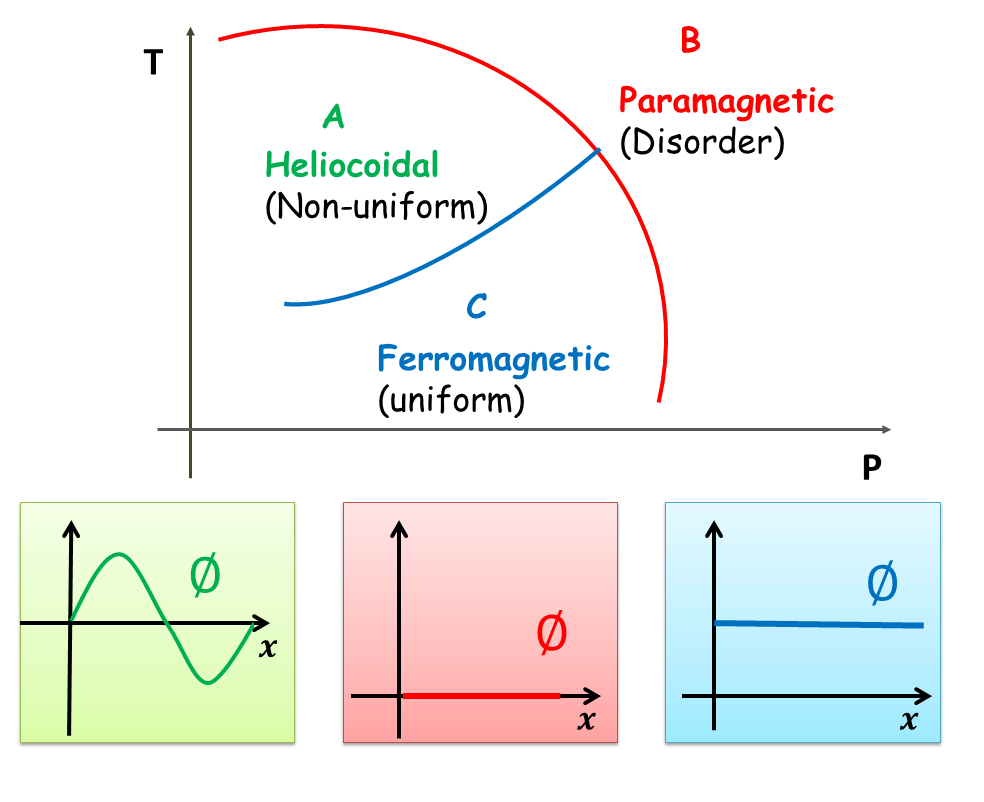}
}
\subfigure[Causal dynamical triangulation.]
{
\includegraphics[width=8.0cm,angle=0]{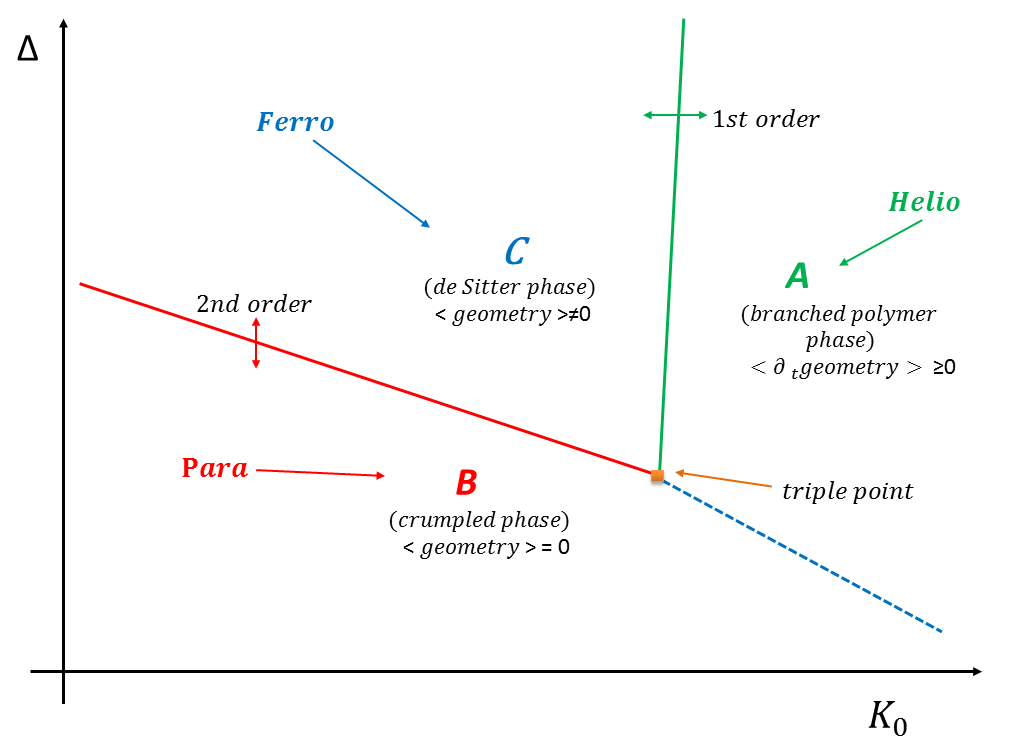}
}
\subfigure[Multitrace matrix model.]
{
\includegraphics[width=12.0cm,angle=0]{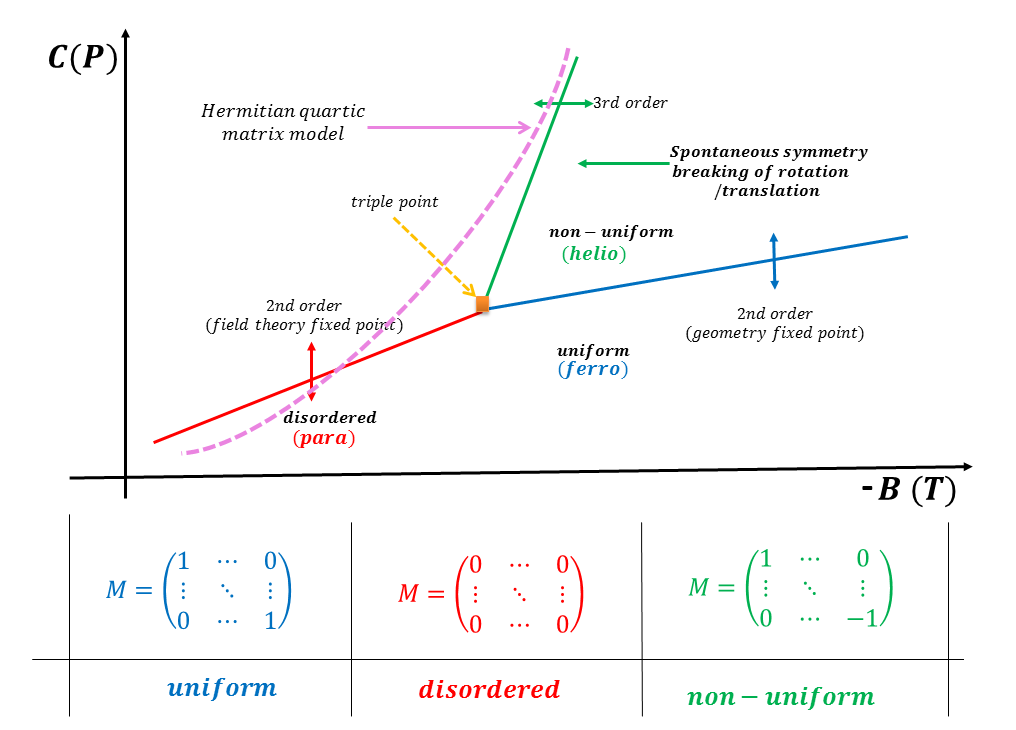}
}
\end{center}
\caption{The phase diagrams of causal dynamical triangulation, Lifshitz scalar field theory and multitrace matrix model.}
\label{CDT}
\end{figure}

Causal dynamical triangulation (CDT) is intimately related to Horava-Lifhsitz (HL) gravity \cite{Horava:2009if,Horava:2008ih,Horava:2009uw} which also, like CDT, assumes global time foliation and introduces anisotropy between space and time but in such a way as to achieve power-counting renormalizability of quantum gravity. This theory is effectively a generalization to gravity of the $d-$dimensional Lifhsitz scalar field theory given by the Lifhsitz-Landau free energy density \cite{Goldenfeld:1992qy}
\begin{eqnarray}
S=a_2\phi^2+a_4\phi^4+...+c_2(\partial_{\alpha}\phi)^2+d_2(\partial_{\beta}\phi)^2+e_2(\partial_{\beta}^2\phi)^2+...\label{LL}
\end{eqnarray}
The anisotropy is introduced by the distinction between the indices  $\beta=1,...,m$ and $\alpha=m+1,...,d$. The three phases present in this theory are: helicoidal ($|\partial_t\phi(x)|<0$), paramagnetic ($\phi(x)=0$) and ferromagnetic ($|\phi(x)|>0$). The phase diagram is depicted in figure $(a)$ of (\ref{CDT}).

The phase structure of causal dynamical triangulation is also summarized in figure $(b)$ of (\ref{CDT}). The cosmological constant $K_4$ which controls the total volume is fixed at its critical value and the phase diagram is then drawn in the plane $K_0-\Delta$ where $K_0$ is proportional to the inverse bare gravitational coupling constant $G$ while $\Delta$ is effectively the asymmetry factor $\alpha$. There are three distinct phases which will be of great importance for our later considerations and thus we describe them in some detail.

We have \cite{Ambjorn:2010hu,Gorlich:2011ga}
\begin{itemize}
\item The de Sitter spacetime phase $C$. This is the analogue of the ferromagnetic phase $(d_2>0, a_2<0$) of the Lifshitz scalar field theory or the ordered phase in noncommutative scalar field theory (see below).
\item The crumpled phase $B$ where neither space nor time have any extent and therefore there is no geometry.  This is the analogue of the paramagnetic phase ($d_2>0, a_2>0$) of the  Lifshitz scalar field theory or the disordered phase in noncommutative scalar field theory (see below). 
\item The branched polymer phase $A$ where the geometry oscillates in time.  This is the analogue of the helicoidal phase ($d_2<0$) in Lifshitz scalar field theory or the non-uniform ordered phase in noncommutative scalar field theory (see below). 
\item The transition from $A$ to $C$ is first order whereas the transition from $B$ to $C$ could be either first order or second order and as a consequence there is a possibility of a continuum limit. Similarly, the transition between the ferromagnetic ($C$) and paramagnetic ($B$) phases in the  Lifshitz scalar field theory although usually second order it could be first order. As we will see a strikingly similar situation occurs in noncommutative scalar field theory (see below). 
\item Also in both theories CDT and HL the spectral dimension at short distances is $2$ and only becomes $4$ at large distances and the anisotropy between space and time disappear in CDT in the de Sitter spacetime phase while in HL it disappears at low energies.
\end{itemize}

\section{The multitrace matrix model}
In this article we wish to propose a new theory of emergent geometry based not on the BFSS and IKKT matrix gauge theories of string theory and noncommutative geometry, i.e. matrix field theory, but based instead on matrix scalar models with a single matrix $M$ enjoying full U(N) symmetry and nothing else. This theory is a generalization of the hermitian quartic matrix model \cite{Brezin:1977sv,Shimamune:1981qf} to  multitrace hermitian matrix models which can be understood as approximations of non-commutative scalar field theory \cite{OConnor:2007ibg}. The corresponding phase structure is very reminiscent to the case found in causal dynamical triangulation and the reason is easily understood in the effective Landau-Lifhsitz scalar field theory (\ref{LL}) which in fact also describes the phase structure of non-commutative scalar field theory \cite{Chen:2001an} and multitrace matrix models \cite{Saemann:2010bw,Nair:2011ux,Tekel:2013vz}. 

We start by writing the star of the show before outlining its derivation, its phase structure and its geometry. We consider a cubic multitrace matrix model of the form
\begin{eqnarray}
V=BTrM^2+CTrM^4+C^{'}TrMTrM^3.\label{cmt}
\end{eqnarray}
This is a three-parameter model which for one range of $C{'}$ (positive values) lies in the universality class of the hermitian  quartic matrix model
\begin{eqnarray}
V_0=BTrM^2+CTrM^4,\label{sc0}
\end{eqnarray}
whereas for another (most important) range of  $C{'}$ (negative values) it lies in the universality class of non-commutative phi-four theory 
\begin{eqnarray}
S=Tr_H(a\Phi\Delta \Phi+b\Phi^2+c\Phi^4).\label{sc}
\end{eqnarray}
The non-commutative scalar phi-four theory (\ref{sc}) is defined on some non-commutative space given by the spectral triple $(A,H,\Delta)$ where $A$ is the algebra of hermitian $N\times N$ matrices ($\Phi\in A$), and $\Delta=[X_a,[X_a,...]]$ is some appropriate Laplacian, where $X_a\in A$ define in a very precise sense the coordinates on the fuzzy space. This theory can be approximated to any arbitrary degree of accuracy with a multitrace matrix model by following the steps: 1) diagonalizing the matrix $\Phi$, 2) expanding the kinetic term in powers of $a$ and 3) performing the U(N) integrals. This is effectively the Hopping parameter expansion on the fuzzy space. We get typically a multitrace action of the form 
\begin{eqnarray}
V=V_0+V_1,
\end{eqnarray}
where
\begin{eqnarray}
V_1&=&E^{'}[TrM^2]^2+B^{'}(TrM)^2+C^{'}TrMTrM^3+D^{'}(TrM)^4+A^{'}TrM^2(TrM)^2+....\nonumber\\\label{sc1}
\end{eqnarray}
The primed coefficients will be different for different fuzzy spaces. For example on the fuzzy sphere we have \cite{Ydri:2014uaa} 
\begin{eqnarray}
E^{'}=3N/4,~~B^{'}=\sqrt{N}/2,~~C^{'}=-N,~~ D^{'}=0, ~~A^{'}=0.
\end{eqnarray}
The phase structure of the above three models (\ref{cmt}), (\ref{sc}) and (\ref{sc0})+(\ref{sc1}) consists typically of the following three phases:
\begin{itemize}
\item The uniform ordered or Ising phase. This is the analogue of the ferromagnetic phase $(d_2>0, a_2<0$) of the Lifshitz scalar field theory. In this case $M\sim {\bf 1}$. The uniform ordered phase is metastable in the hermitian quartic matrix model $V_0$ \cite{Brezin:1977sv,Shimamune:1981qf}.
\item The disordered phase. This is the analogue of the paramagnetic phase ($d_2>0, a_2>0$) of the  Lifshitz scalar field theory. In this case $M\sim 0$.
\item The non-uniform ordered or stripe phase. This is the analogue of the helicoidal phase ($d_2<0$) in Lifshitz scalar field theory. In this phase $M\sim \gamma$ where $\gamma^2=1$. Thus, in this phase translational/rotational invariance is spontaneously broken.
\item The most important transition for emergent geometry is the commutative space transition from uniform to disordered which is second order and  hence defines a field theory continuum limit (see below). Whereas the transition from uniform  to non-uniform could be first order but more likely second order \cite{Gubser:2000cd,Ambjorn:2002nj} since it is the continuation of the Ising line and hence there is the possibility of obtaining a geometry continuum limit here. Finally, the transition between the non-uniform to disordered is the hermitian quartic matrix model third order phase transition.
\end{itemize}
The phase diagram is sketched in figure $(c)$ of (\ref{CDT}). We can immediately notice the striking similarities between the three phase diagrams of causal dynamical triangulation, non-commutative scalar phi-four and Landau-Lifshitz scalar field theory. However, as opposed to dynamical triangulation where the spacetime emerges only in the de Sitter space phase $C$, in our case the geometry of space is well established in two phases: the disordered and the uniform phases and the transition between them as we vary the temperature, played here by the parameter $B$, is the ordinary field theory continuum limit (the famous Ising transition) on that space. 

Also as opposed to dynamical triangulation if we vary the temperature for a value of the pressure played here by the parameter $C$ above the triple point the geometry evaporates in a third order matrix transition between disordered and non-uniform, where rotational invariance gets spontaneously broken (note the similarity with the case of the Lorentzian IKKT model), then as we increase the temperature further the geometry emerges again in a second order phase transition between non-uniform and uniform phase. This transition can be thought of as a geometry continuum limit similarly to what happens in dynamical triangulation between phases $C$ and $B$.

This picture is expected to hold in any dimension on any fuzzy space and the three phases meet at a triple point. But it is also expected to hold in particular multitrace matrix models. See \cite{Ydri:2015vba} and extensive list of references therein for the actual Monte Carlo calculation of the phase diagrams on the fuzzy sphere and of various multitrace approximations thereof. As we have already mentioned the uniform ordered phase is metastable in the lowest order approximation given by the hermitian quartic matrix model but becomes stable in the full non-commutative model and thus there should exist a minimal truncation of the full multitrace matrix model which exhibits this uniform phase on which  the geometry can be consitently defined. Indeed, it was remarked in \cite{Ydri:2015vba}  that the cubic term $TrMTrM^3$  is the only term we need to add to the hermitian quartic matrix model $V_0$ with the desired effect of producing a phase diagram containing a uniform phase. The cubic multitrace matrix model (\ref{cmt}) is therefore the minimal truncation of the noncommutative phi-four theory with an emergent fuzzy sphere as we will further describe below.

The central proposal of this article is to turn the original logic on its head and regard the multitrace matrix model (\ref{sc0})+(\ref{sc1}) as a starting point, i.e. as a first principle, and then try to determine the geometry of the space from the properties of the uniform-to-disordered phase transition if present. The potential (\ref{sc0})+(\ref{sc1}) is a matrix model of a single hermitian matrix $M$ with unitary U(N) invariance, i.e. without a kinetic term, and thus no geometry a priori. We arrive then at our first rule:
\[{I: \it The~uniform~second~order~phase~transition}\Rightarrow {\it geometry}.\]
The dimension of this space is determined from the critical exponents of the uniform-to-disordered phase transition by virtue of scaling and universality of second order phase transitions. We have then the second rule:
\[{II: \it Critical~exponents}\Rightarrow {\it dimension}.\]
The metric on the space is encoded in the free propagator which in turn is encoded in the eigenvalue distribution of the matrix $M$ which must follow a Wigner semicircle law with a particular radius depending crucially on the kinetic term. We have then the third rule:
 \[{III: \it Wigner~semicircle~law}\Rightarrow {\it metric}.\]

Each fuzzy geometry will correspond to a set of primed coefficients. We divide these coefficients into three sets: 1) $\alpha=C^{'}$ , 2) $\beta=\{E^{\prime},...\}$, 3) $\gamma=\{B^{\prime},D^{\prime},A^{\prime},...\}$. The second set consists of operators which depend only on even moments whereas the third set consists of operators which involve odd moments. The first set consists only of the term $TrMTrM^3$ which is sufficient on its own as we will see to produce the most fundamental of all geometries, i.e. the sphere. Recall that the moments of the matrix $M$ are defined by $m_n={Tr}M^n$.

Then every geometry which admits a finite spectral triple corresponds to a fixed point in the infinite dimensional space $\alpha-\beta-\gamma$. For example, at the origin of this space we find the hermitian quartic matrix model. On the other hand, not all terms in a given multitrace expansion on some fuzzy space are necessary to produce the geometry of that particular fuzzy space. Indeed, a truncation of this expansion may be sufficient. For example,  the matrix model  (\ref{cmt}) is one such truncation (in fact the minimal one) of the multitrace expansion which will produce the geometry of the sphere  ${\bf S}^2$. 

\begin{figure}[H]
\begin{center}
\includegraphics[angle=-0,scale=0.6]{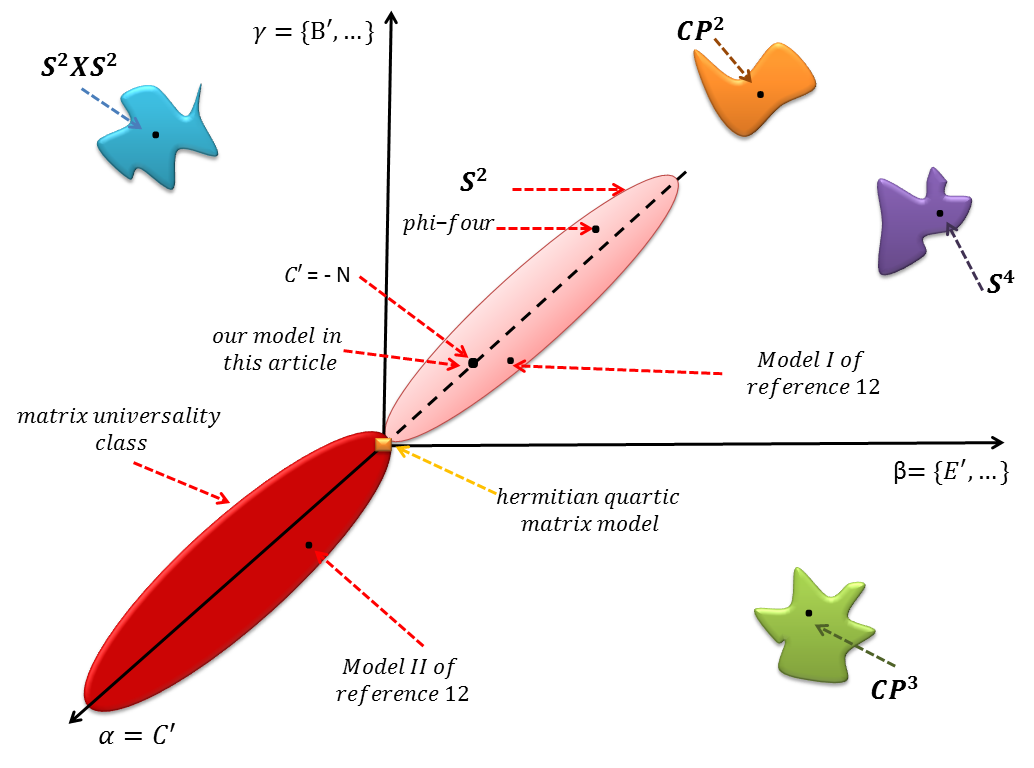}
\end{center}
\caption{The phase structure of the multitrace matrix model in the  $\alpha-\beta-\gamma$ space.}\label{sketch}
\end{figure}

Other geometries will obviously require the addition of one or more of the other multitrace operators. Hence, we can also have all fuzzy spaces such as fuzzy ${\bf S}^2\times {\bf S}^2$, fuzzy ${\bf S}^4$, fuzzy ${\bf CP}^n$  and in general any other fuzzy or noncommutative space which can be defined by a finite spectral triple. However, in the space $\alpha-\beta-\gamma$ there are distinct whole regions for different spaces, since the truncation of the multitrace expansion is not unique, and these regions may be connected by  various paths via varying one or more of the primed coefficients. See figure (\ref{sketch}). The geometry in each region emerges dynamically in the quantum theory defined by the partition function 

\begin{eqnarray}
Z&=&\int {\cal D}M~ \exp(-V),
\end{eqnarray}
where
\begin{eqnarray}
V&=&BTrM^2+CTrM^4+\alpha TrMTrM^3+\beta_1 (TrM^2)^2+\gamma_1(TrM)^2+...\label{our_model}
\end{eqnarray}
This potential should be thought of as a generalization of the discretization of random Riemannian surfaces (two-dimensional quantum gravity or $D=0$ bosonic string theory) with regular polygons with $j>1$ vertices (generalization of dynamical triangulation) given by \cite{DiFrancesco:1993cyw,Zarembo:1998uk}
\begin{eqnarray}
V&=&B_2TrM^2+B_3 TrM^3+ B_4 TrM^4+....\label{mo}
\end{eqnarray}
In summary, we can have in the case of the multitrace matrix models emergent geometry as well as growing dimensions and topology change. More precisely, the model (\ref{our_model}) seems to work in two dimensions (as the model (\ref{mo})) but also away from two dimensions where it captures a large class of spaces which admit finite spectral triples.

\section{The non-pertrubative phase diagram}
We now discuss the case of the minimal truncation (\ref{cmt}) in more detail. We can apply all non-pertrubative methods at our disposal: Monte Carlo simulation, random matrix theory and renormalization group where $1/N$ plays a major role in all three. 

\begin{figure}[H]
\begin{center}
\subfigure[The phase diagram where $\tilde{B}=BN^{-3/2}$ and $\tilde{C}=CN^{-2}$. The triple point is ($\tilde{B},\tilde{C}$)=($-3.1$ ,$ 1.95$).]
{
\includegraphics[angle=-0,scale=0.4]{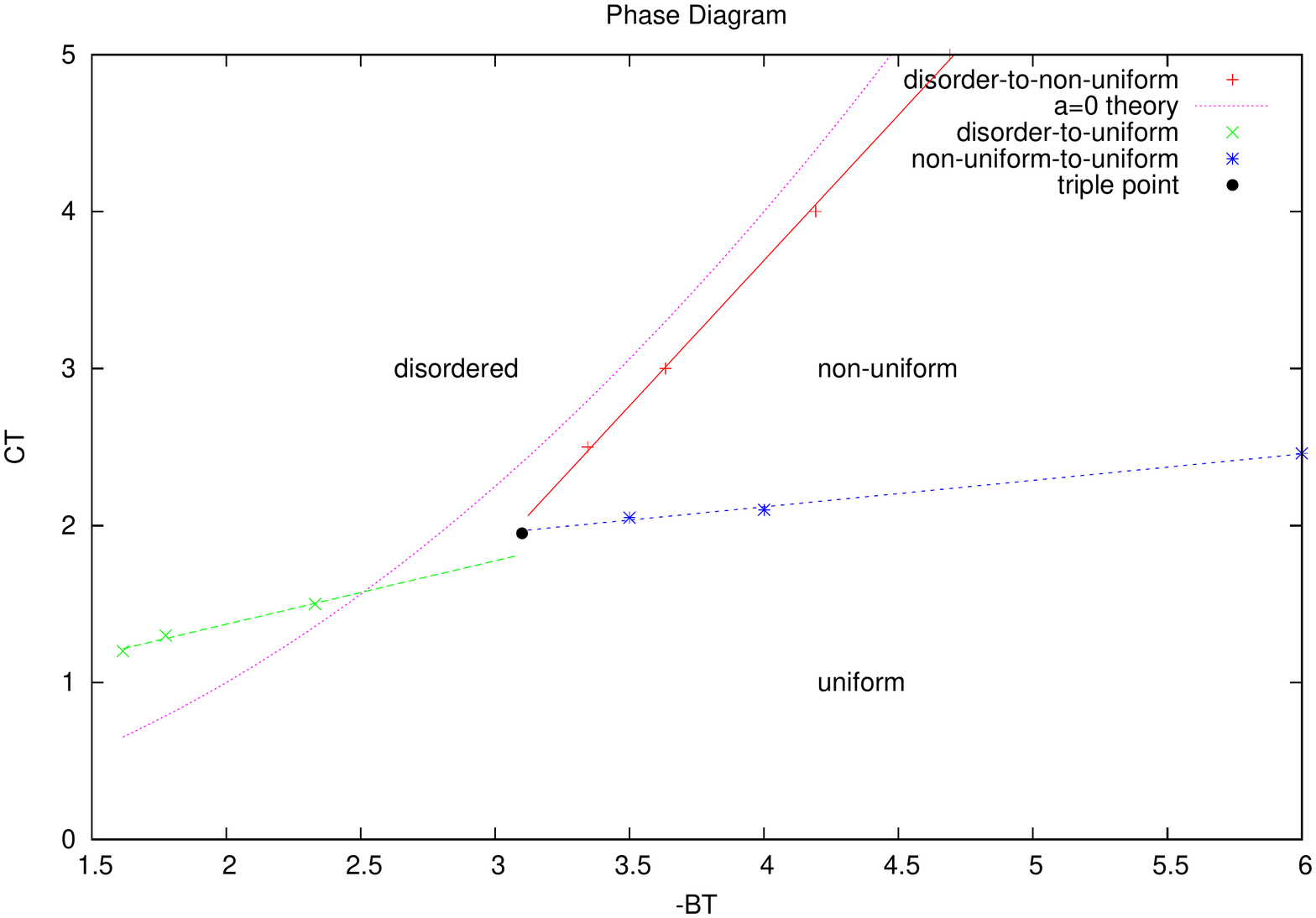}
}
\subfigure[The critical exponents $\beta$, $\gamma$, and $\alpha$.]
{
\includegraphics[angle=-0,scale=0.4]{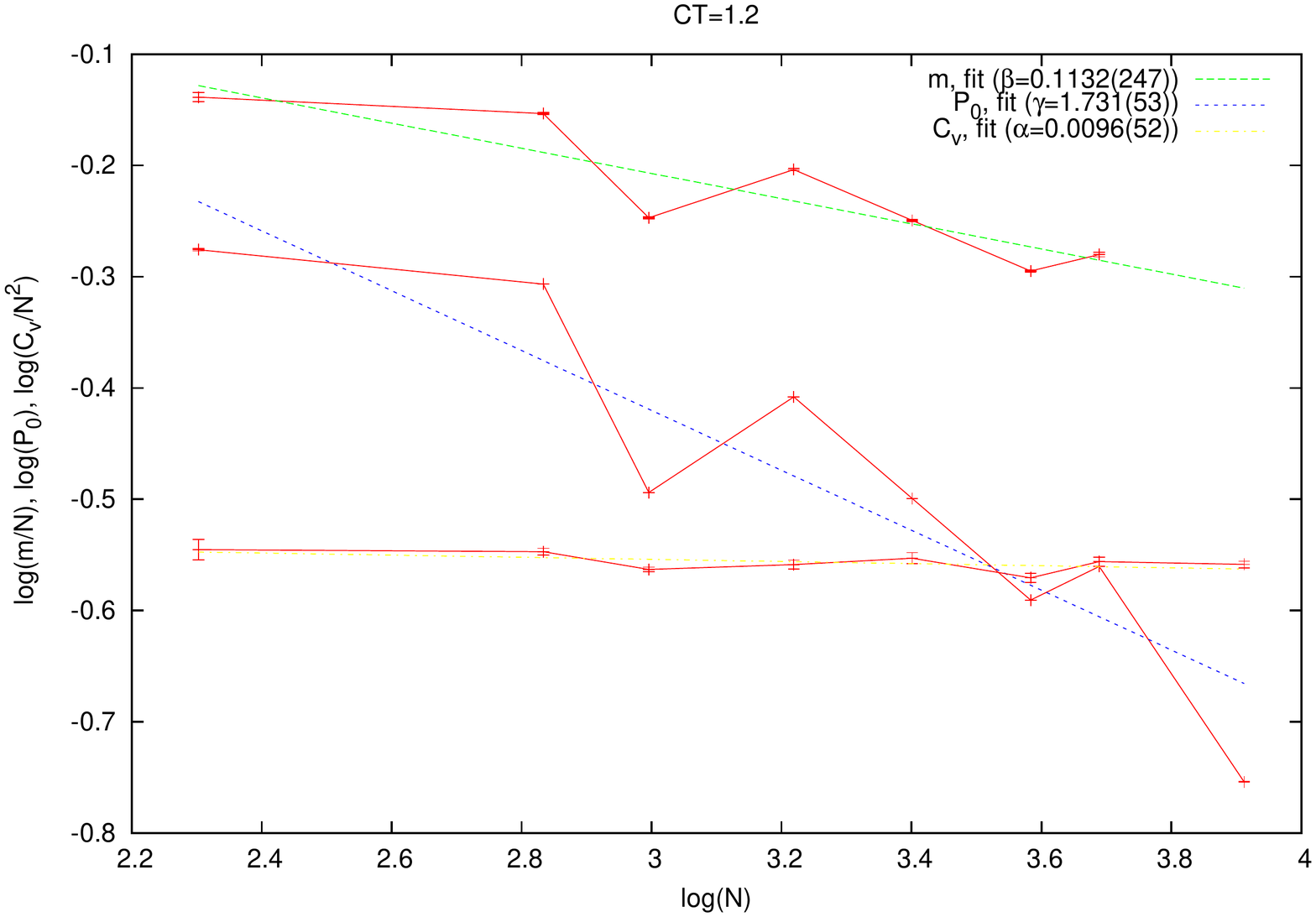}
}
\caption{The phase diagram of the of the minimal truncation (\ref{cmt}) and its critical exponents.}
\label{pds}
\end{center}
\end{figure}

The cubic term  $TrMTrM^3$ with any negative coefficient $C^{'}$ added to the hermitian quartic matrix model $V_0$ is sufficient to generate a stable uniform phase on the fuzzy sphere. The actual non-perturbative phase diagram obtained by means of Monte Carlo simulation for $C^{'}=-N$ is shown on figure (a) of (\ref{pds}). 

The critical exponents are found to be consistent  with the Onsager values in two dimensions \cite{Onsager:1943jn1} given by  $\nu=1,\beta=1/8,\gamma=7/4,\alpha=0,\eta=1/4,\delta=15$. For example, the critical exponent $\beta$ associated with the behavior of the magnetization $m=\langle |{Tr}M|\rangle$ near the critical temperature  is found to be given by $\beta=0.1132(247)$. Measurement of the magnetization $m/N$ as a function of the temperature $B_T=\tilde{B}=BN^{-3/2}$ for $C_T=\tilde{C}=C/N^2$ at the discontinuity is shown on figure (b) of (\ref{pds}). Also the measurement of the zero power $P_0=\langle \big({Tr} M/N\big)^2\rangle$ (related to susceptibility) and the specific heat which yield the critical exponents $\gamma$ and $\alpha$ (the exponent $\nu$ is determined by scaling) is shown on figure (b) of (\ref{pds}).

\begin{figure}[H]
\begin{center}
\subfigure[The Wigner semicircle law.]
{
\includegraphics[width=8.0cm,angle=-0]{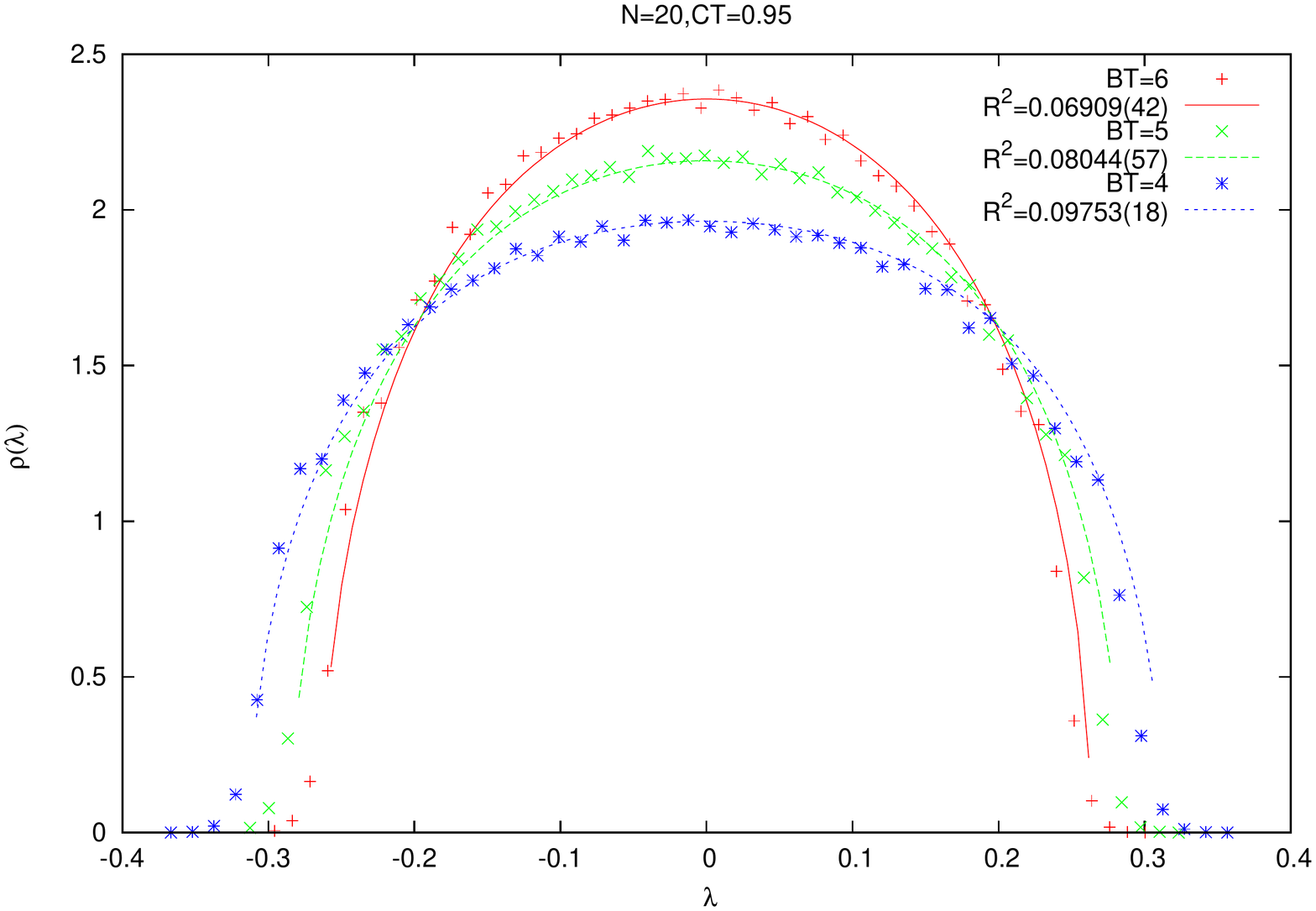}
\includegraphics[width=8.cm,angle=-0]{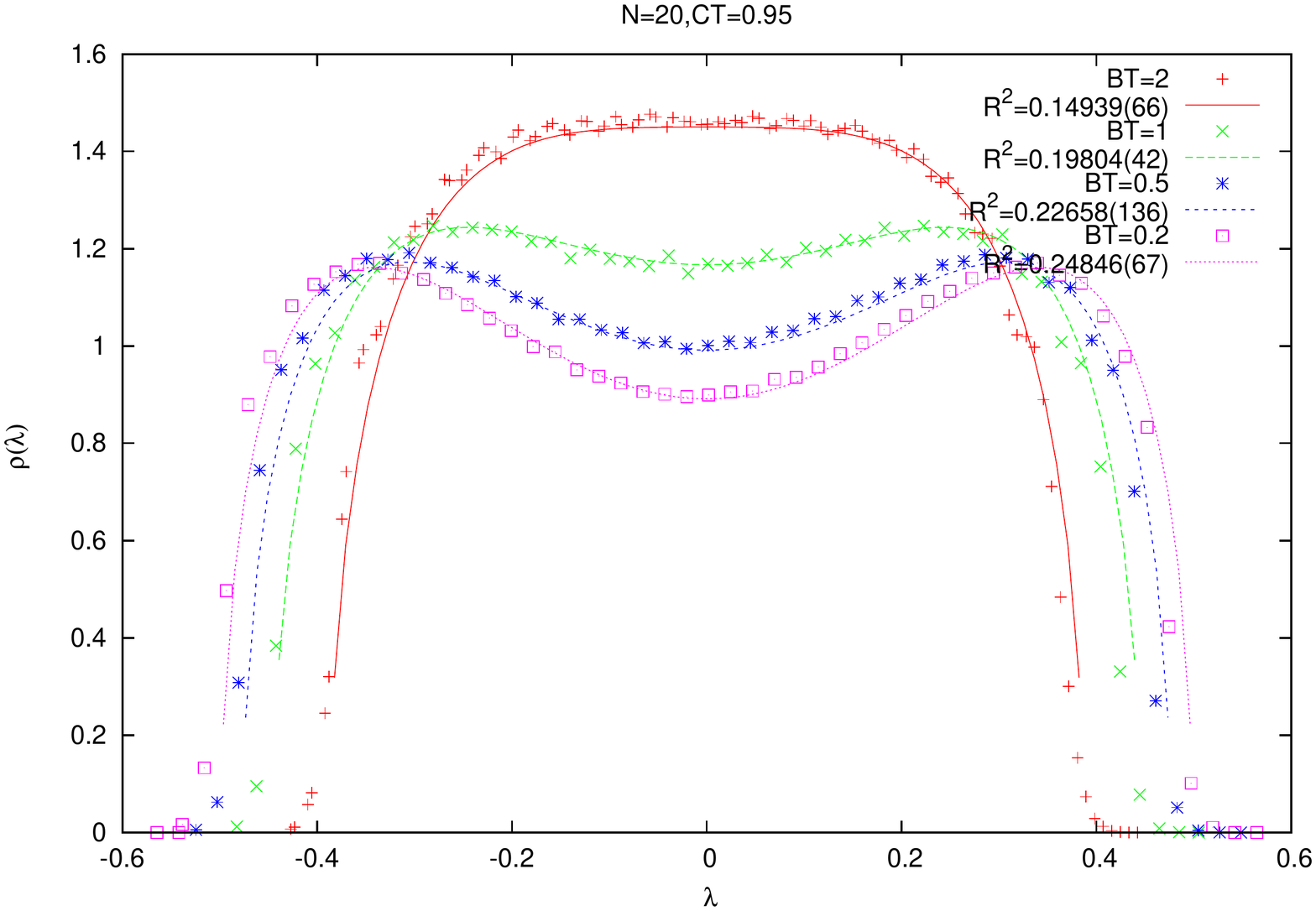}
}
\subfigure[The radius (largest eigenvalue) as a function of the mass.]
{
\includegraphics[width=8.0cm,angle=-0]{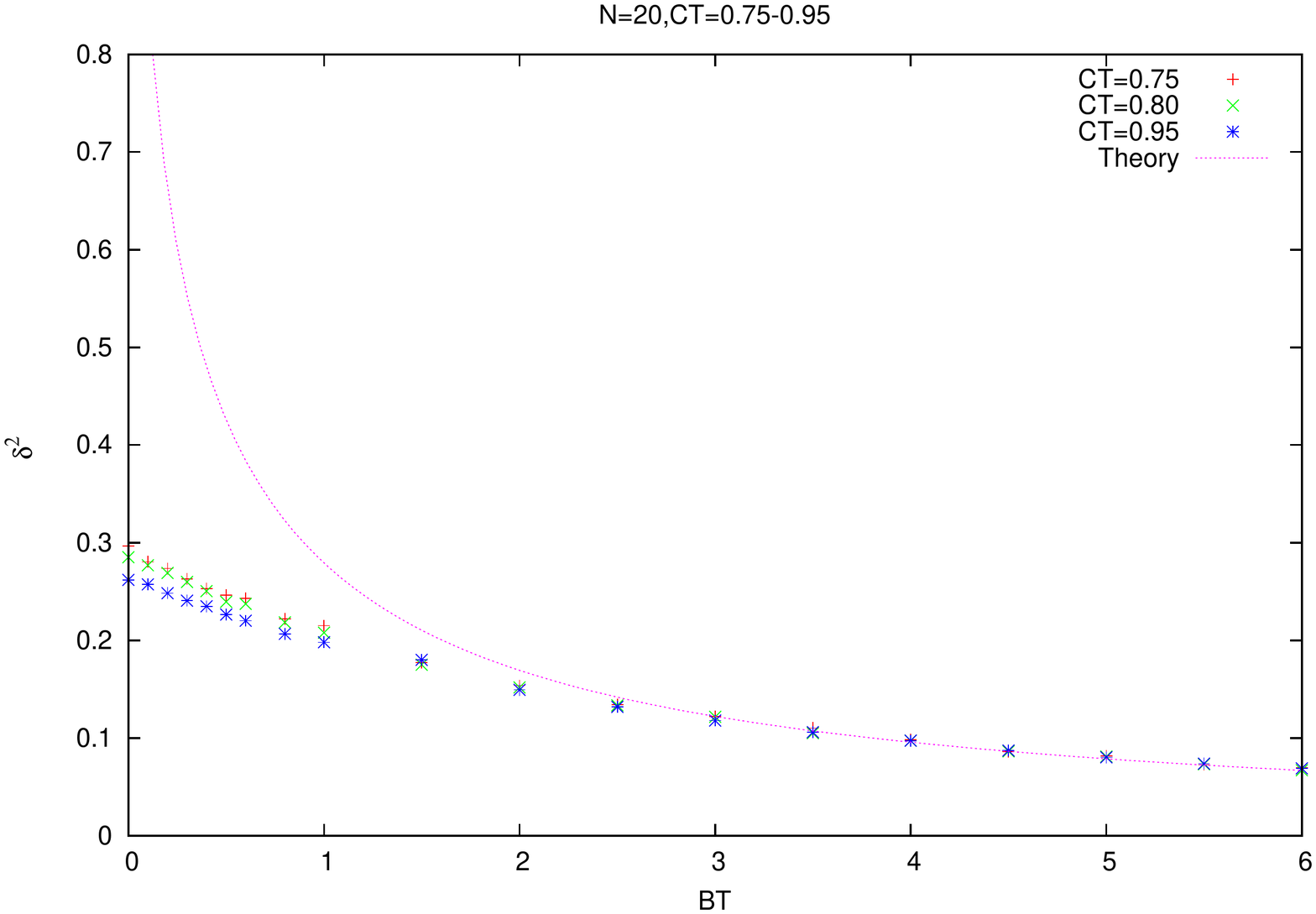}
}
\caption{Wigner semicircle law.}
\label{WSC}
\end{center}
\end{figure}

In order to discuss the measurement of the free propagator from the Wigner semicircle law we recall few results. A non-commutative free scalar field theory in $d$ dimensions in the matrix basis is an ${\cal N}\times {\cal N}$ matrix model which may be defined by means of a sharp cutoff $\Lambda$.  Since planar diagrams dominates over the non-planar ones in the limit $\Lambda\longrightarrow \infty$ the probability distribution of the eigenvalues $\Phi_i$ of the free scalar field $\Phi$ can be shown \cite{Steinacker:2005wj,Nair:2011ux} to be given by a Wigner semicircle law with radius $R$ given by the largest eigenvalue $\alpha_0$ which depends on the dimension $d$ and the cutoff $\Lambda$.  By hindsight we know that the cubic term  $TrMTrM^3$ alone can only capture two dimensions on the fuzzy sphere, i.e. ${\cal N}=N+1$ and $\Lambda=N/{\cal R}$ where ${\cal R}$ is the radius of the sphere, and thus we have a prediction for $\alpha_0$ given by \cite{Steinacker:2005wj}
\begin{eqnarray}
\alpha_0^2=\frac{1}{\pi}\ln \bigg(1+\frac{2\pi}{\sqrt{N}\tilde{B}}\bigg).
\end{eqnarray}
The actual Monte Carlo measurement of the radii $R=\alpha_0$ for various values of $\tilde{B}$ is obtained by fitting the data to the Wigner semicircle law and then it is plotted and compared with the above prediction. See figure (\ref{WSC}). The agreement is very good with some deviation as we approach the non-perturbative region where the uniform ordered phase appears at some $\tilde{B}<0$. 

However, there is also another source of discrepancy which lies in the fact that the above results for the behavior of the propagator are with the reference to the free theory $C=0$. On the other hand, we can not reach the limit $C\longrightarrow 0$ since the model  (\ref{cmt}) is stable, i.e. the quartic term $CTr M^4+C{\prime}TrMTrM^3$ is positive definite (substitute for example $M\sim {\bf 1}$), only for  $\tilde{C}\geq 1$.

The above picture can also be confirmed by solving in $1/N$ the eigenvalue problem in the various phases. The phase structure is found to contain a  stable uniform ordered phase and the triple point can be estimated. See for example chapter $5$ of \cite{Ydri:2016dmy} for the detailed solution of a related general class of multitrace matrix models.

A much more powerful and illuminating approach is the matrix renormalization group approach of \cite{Higuchi:1994rv}.  Recall that $\tilde{B}=BN^{-3/2}$, $\tilde{C}=C/N^2$ and similarly we will define $\tilde{C}^{\prime}=C^{\prime}/N$. The model (\ref{cmt}) is defined only for $\tilde{C}>-\tilde{C}^{\prime}$ and all phase transitions occur for $\tilde{B}<0$. If we analytically continue the model to the values $\tilde{C}<-\tilde{C}^{\prime}$ (see the explicit and pedagogical discussion found in \cite{Marino:2015}), then we can assume that $\tilde{B}>0$. This can also be seen by scaling the field appropriately to bring the potential to the form (with $\mu=B/|B|$, $g_3=3\tilde{C}^{\prime} a/4\tilde{B}^2$, $g_4=\tilde{C}/\tilde{B}^2$)
\begin{eqnarray}
V=N\bigg(\frac{\mu}{2}TrM^2+\frac{g_3m_1}{3a}TrM^3+\frac{g_4}{4}TrM^4\bigg).
\end{eqnarray}
The $m_1$ is the first moment $m_1=TrM/N$ and $a$ is its classical value, viz the solution of the equation of motion $\mu\phi+g_4\phi^3+g_3m_1\phi^2/a+g_3m_3/3a=0$. We find explicitly 
\begin{eqnarray}
a=0~,~a=\sqrt{-\frac{\mu}{\tilde{C}+\tilde{C}^{\prime}}}|\tilde{B}|.
\end{eqnarray}
The first solution is the disordered value whereas the second is the uniform value which makes sense only if $\mu<0$,$\tilde{C}>-\tilde{C}^{\prime}$ or $\mu>0$, $\tilde{C}<-\tilde{C}^{\prime}$. 

To simplify the discussion here we may also employ the mean field approximation in which we replace the first moment by its average, viz $TrM=Na+...$ and thus $g_3$ becomes a constant. Then, the solution of the matrix renormalization group equation consists of $4$ fixed points $(g_{3*},g_{4*})$ \cite{Higuchi:1994rv}
\begin{itemize}
\item The hermitian quartic matrix model fixed points:
\begin{itemize}
\item A Gaussian fixed  point $(0,0)$ related to the disordered phase.
\item A $(2,3)$ conformal quantum gravity fixed point $(0,-1/12)$ related to the third order transition to the non-uniform phase. 
\end{itemize}
\item The multitrace matrix model fixed points relevant to the operator $TrMTrM^3$:
\begin{itemize}
\item A second $(2,3)$ conformal quantum gravity fixed  point $(0.22,0)$.
\item A $(2,5)$ conformal quantum gravity fixed point $(0.31,0.03)$. 
\end{itemize}
These points seems to be related to the second order phase transitions disordered-to-uniform and non-uniform-to-uniform.
\end{itemize}
The precise relation between the above structure of fixed points and the phase structure of the multitrace matrix model will be discussed elsewhere together with corrections away from the mean field approximation used above \cite{in_preparation}.

\section{Emergent gauge theory and emergent gravity}
The emergent geometry can be exhibited  in a drastic way by exhibiting an emergent non-commutative gauge theory on a fuzzy sphere in the uniform phase \cite{Ydri:2016daf}. We will assume now that the matrix $M$ is $2N\times 2N$. Then, without any loss of generality, we can expand the matrix $M$ as
\begin{eqnarray}
M=M_0{\bf 1}_{2N}+M_1~,~Tr M_1=0.
\end{eqnarray}
Hence
\begin{eqnarray}
M_1=\sigma_aX_a~,~
M_0=a+m,
\end{eqnarray}
where $\sigma_a$ are the standard Pauli matrices, $m$ is the fluctuation in the zero mode, and $X_a$ are three hermitian $N\times N$ matrices. By substitution, we obtain immediately the model
\begin{eqnarray}
Z&=&\int {\cal D} X_a \exp(-V[X]))\int dm\exp(-f[m]).
\end{eqnarray}
The potential $V$ is given now by the $SO(3)-$symmetric three matrix model 
\begin{eqnarray}
V&=&-CTr[X_a,X_b]^2+2CTr(X_a^2)^2+2(B+6Na^2C^{\prime})TrX_a^2+4iN aC^{\prime}\epsilon_{abc} TrX_aX_bX_c.\nonumber\\
\end{eqnarray} 
The integration over $m$ can be done and the result consists of some function of $TrX_a^2$ and $i\epsilon_{abc}TrX_aX_bX_c$. This next to leading contribution (in $1/N$)  is essentially the one-loop result and it is by construction subleading compared to $V$.

The Chern-Simons term is proportional to the value $a$ of the order parameter. Thus, it is non-zero only in the Ising phase, and as a consequence, by tuning the parameters appropriately to the region in the phase diagram where the Ising phase exists, we will induce a non-zero value for the Chern-Simons. This is effectively the Myers term responsible for the condensation of the geometry \cite{Myers:1999ps,Azuma:2004zq}. The above multitrace three matrix model is then precisely a random matrix theory describing non-commutative gauge theory on the fuzzy sphere, where the first term is the Yang-Mills piece, whereas the second and third terms combine to give mass and linear terms for the normal scalar field on the sphere (recall that the index $a$ runs from $1$ to $3$). This is essentially the random matrix theory describing non-commutative gauge theory on the fuzzy sphere  found in   \cite{Steinacker:2003sd}. However, we should emphasis here that we have obtained dynamically this gauge theory on the fuzzy sphere by going to the phase where a non-zero uniform order persists, and then by expanding around this order, thus securing a non-zero Chern-Simons term crucial for the condensation of the fuzzy sphere geometry. In \cite{Steinacker:2003sd}, this was achieved by constraining the matrix $M$ directly by hand in a particular way.

By expanding the above gauge theory around the background solution (a fuzzy sphere solution), then employing the Weyl map and the star product on the fuzzy sphere \cite{Presnajder:1999ky}, we obtain a non-commutative U(1) gauge theory on the sphere with pointwise multiplication replaced with the star product. By  taking then the planar limit of  \cite{Alexanian:2000uz}, we obtain a U(1) gauge theory on the non-commutative plane ${\bf R}^2_{\theta}$. However, we know that non-commutative U(1) gauge theory is effectively a gravity theory. Indeed, by using the Seiberg-Witten map \cite{Seiberg:1999vs} we can determine the dual emergent gravity of this non-commutative U(1) gauge theory \cite{Rivelles:2002ez} which turns out to be a pp-wave spacetime \cite{final_ref}. 

A more systematic approach to emergent gravity in matrix models (in particular the IKKT matrix model) can be found in \cite{Steinacker:2010rh,Yang:2008fb}.

\section{Conclusion}
The above discussion can be extended virtually as it stands to Cartesian products of fuzzy spheres such as ${\bf S}^2\times{\bf S}^2$, etc. The criterion for quantum gravity as emergent discretized random spaces given by the suppositions I, II and III (page $8$) can also be applied to fuzzy ${\bf S}^4$ and in fact to any fuzzy space which admit a finite spectral triple. However, exhibiting explicitly non-commutative gauge theories and emergent gravity on these fuzzy spaces by expanding around the uniform order is expected to be quite difficult. Extension to Lorentzian signature seems also to be within reach \cite{Chaney:2015mfa}.

\paragraph{Acknowledgments:}
This research was supported by CNEPRU: ``The National (Algerian) Commission for the Evaluation of University Research Projects''  under contract number ${\rm DO} 11 20 13 00 09$. B.Y would like also to acknowledge funding from the Abdus Salam International Center for Theoretical Physics (ICTP) under the  ``Associate Scheme''. Also, he would like to thank the Dublin Institute for Advanced Studies for warm hospitality on numerous occasions as a ``Research Associate''.

\end{document}